\documentclass[11pt,fleqn,a4paper]{article}

\usepackage{titling}
\usepackage{preliminary}
\usepackage{hyphenat}
\usepackage[authoryear]{natbib}
\usepackage{amssymb}
\usepackage{amsthm}
\usepackage{epstopdf}
\usepackage{amsmath}

\usepackage{scrlayer-scrpage}
\usepackage{subfigure}
\usepackage{setspace}
\usepackage[bottom]{footmisc}
\usepackage{endnotes}
\usepackage{graphicx}
\usepackage{textcomp}
\usepackage{natbib}
\usepackage[utf8]{inputenc}
\usepackage{verbatim}
\usepackage{lscape}
\usepackage{array}
\usepackage{fltpoint}
\usepackage{dcolumn}
\usepackage{booktabs}
\usepackage{rotating}
\usepackage[norounding,point]{rccol}
\usepackage{hyperref}
\usepackage{placeins}
\usepackage{multirow}
\usepackage{paralist}
\usepackage{enumitem}
\usepackage{textgreek}
\usepackage{float}
\usepackage{graphicx}
\graphicspath{ {./Images/} } 
\usepackage{tablefootnote}
\usepackage{eurosym}

\usepackage{multirow}

\usepackage{tikz-cd}
\usepackage{ctable}

\usepackage{tikz}
\usetikzlibrary{positioning}
\usetikzlibrary{snakes}

\usepackage[toc,page]{appendix}

\usepackage{multicol}
\usepackage{caption}

\usepackage{subfigure}

\usepackage{color}

\usepackage{amsmath}
\usepackage{url}

\usepackage{geometry}
\numberwithin{equation}{section}

\hypersetup{
colorlinks=true,
urlcolor=blue,
linkcolor=red,
citecolor=blue
}
\defcitealias{AUSSENHANDELSSTATISTIK2018}{AS, 2018a}
\defcitealias{Volksw}{AS, 2018d}
\defcitealias{AFS2020}{AS, 2020}
\defcitealias{Beschaftigungsstatistik2019}{AS, 2019}
\defcitealias{Beschaftigungsstatistik2018}{AS, 2018b}
\defcitealias{Lohnstatistik2018}{AS, 2018c}
\defcitealias{BFS2020}{BFS, 2020}

\setheadwidth{textwithmarginpar}
\usepackage{setspace}
\usepackage{xspace}

\numberwithin{theorem}{section}
\numberwithin{assumption}{section}

\newcommand{\auslassen}[1]{}

\usepackage{graphicx}

\usepackage{chngcntr}
\counterwithin{figure}{section}
\counterwithin{table}{section}

\usepackage[nottoc]{tocbibind}

\usepackage{blindtext}
\newgeometry{top=3cm,bottom=3cm,right=2.5cm,left=2.5cm}

\begin{document}

\begin{center}
{\LARGE Do soda taxes affect the consumption and health of school-aged children? Evidence from France and Hungary$^*$}\label{s:soda}

\large \vspace{0.8cm}

{\large  Selina Gangl$^\dagger$}\medskip

{\small {$^\dagger$University of Fribourg (Switzerland), Department of Economics} }\\

\today

\end{center}

\begin{abstract}
This paper examines the effect of two different soda taxes on consumption behaviour and health of school-aged children in Europe: Hungary imposed  a Public Health Product Tax (PHPT)  on several unhealthy products in 2011. France introduced solely a soda tax, containing sugar or artificial sweeteners, in 2012. In order to exploit spatial variation, I use a semi-parametric Difference-in-Differences (DID) approach. Since the policies differ in Hungary and France, I analyse the effects separately by using a neighbouring country without a soda tax as a control group.  The results suggest a counter-intuitive positive effect of the tax on soda consumption in Hungary. The reason for this finding could be the substitution of other unhealthy beverages, which are taxed at a higher rate, by sodas. The effect of the soda tax in France is as expected negative, but insignificant which might be caused by a low tax rate. The body mass index (BMI) is not affected by the tax in any country. Consequently, policy makers should think carefully about the design and the tax rate before implementing a soda tax. \\ 
\\[0.05cm]

\end{abstract}
\vfill
\begin{singlespace}
{\scriptsize 
$^*$ I have benefited from comments by seminar participants at the Swiss Public Health Conference 2020,  the EuHEA PhD Student-Supervisor \& Early Career Researcher Conference 2020, the Annual Congress of the Verein f\"ur Socialpolitik 2020, and a research seminar at the University of Fribourg. I thank participants, in particular Edel Doherty,  Reiner Eichenberger, Martin Huber,  Marco Portmann, Giannina Vaccaro, Christian Zihlmann,  and two anonymous reviewers of SMYE 2021 for their helpful comments.  Addresses for correspondence:  Selina Gangl, University of Fribourg, Bd.\ de P\'{e}rolles 90, 1700 Fribourg, Switzerland;  selina.gangl@unifr.ch. Declaration of conflicts of interest: none. Corresponding author: Selina Gangl. \\
}
\end{singlespace}
{\small \renewcommand{\thefootnote}{\arabic{footnote}} %
\setcounter{footnote}{0}  \pagebreak \setcounter{footnote}{0} \pagebreak %
\setcounter{page}{1} }
\pagestyle{plain}

\pagenumbering{arabic}

\pagestyle{plain}
\newgeometry{top=2cm,bottom=4cm,right=2cm,left=2cm}

\newpage

\section{Introduction}\label{s:intro}
Childhood obesity is a worldwide problem and implies risks like staying obese in adulthood as well as developing non-communicable diseases \citep{WHO2020}. Therefore,  ways and means are searched to mitigate or even prevent childhood obesity. Since an association between childhood obesity and soda consumption exists,  the intake of sugar-sweetened drinks among children should be reduced \citep{JamesKerr2005}.
For example,  a substitution of sugar-sweetened drinks by sugar-free drinks showed a decrease in weight among children \citep{Ruyter2012}.

Hungary and France have implemented a tax on sodas, yet the design and the tax rate differed among the countries: Hungary introduced  a broad tax on unhealthy products, in which sugar-sweetened sodas were taxed by converted 1.83 Eurocents per litre\footnote{converted into Euros at the rate on 01.09.2011} from 2011 on \citep{Ecorys2014}. One year later, France imposed a soda tax of 7.16 Eurocents per litre but included sodas with artificial sweetener too \citep{Ecorys2014}.  These taxes are designed to increase the price  and aim to decrease the consumption of sodas in the population. The existing literature focuses on the consumption behaviour of the household, whereas little is known about the response of children to a soda tax \citep{NZIER2017}.

Hence, this paper analyses the effect of a  soda tax on consumption behaviour and the body mass index (BMI) of school-aged children in Hungary  and France.  Since the policies differ  among the countries, I analyse the effect separately. In the first analysis, Hungary forms the treatment group, while neighbouring country Croatia, which does not levy a soda tax, serves as the control group. In the second analysis, France constitutes the treatment group and Spain, the control group without a soda tax. Methodologically, I exploit spatial variation of the tax and use a semi-parametric Difference-in-Differences (DiD) approach to evaluate the policy. This method uses inverse probability weighting (IPW) to control for differences in observable characteristics between the treatment and control group as well as over time.

Since this paper aims at a cross-country comparison, I use data from the cross-national survey Health-Behaviour in School-Aged Children (HBSC) which ensures that the same question is asked in each country. This survey is conducted in cooperation with the World Health Organization (WHO) Europe and takes place on a quadrennial basis. In the setting of this natural experiment, the year 2010 constitutes the pre-treatment period and 2014 post-treatment period. Furthermore, I use the survey years 2006 and 2010 as pre-treatment years to provide evidence for the parallel trend assumption. 

The results suggest a counter-intuitive positive and significant effect of the tax on soda consumption in Hungary. Considering the nature of the tax might explain this finding. Since the prices of other unhealthy products, like energydrinks, increase as well, the substitution behaviour of children could explain this result. Soda consumption is not affected by the tax in France. A reason for this finding might be the low soda tax rate of 7 Eurocents per litre. Children's body mass index is not influenced by the tax in any country.

This paper relates to two strands of literature. The first one addresses the impact of a soda tax on adults or households in different countries and several U.S. jurisdictions. For example, \cite{Falbe2016} focus on poorer districts in North California and study the effect of a converted 31 Eurocents per litre\footnote{converted into  litre and Euros (at the rate on 01.03.2015)}  soda tax in Berkeley in 2015, compared to similar areas without such a tax in Oakland and San Francisco. They find a decrease in soda consumption  by 21\% after the tax implementation and an increase in water consumption by 63\%. In Philadelphia, the soda tax amounts to  converted 49 Eurocents per litre\footnote{converted into litre and Euros (at the rate on 01.01.2017)} in 2017. \cite{Zhong2018} find that  Philadelphians, compared to citizens of close and similar cities without such a tax,  drink 40\% fewer sodas directly after the tax implementation, yet the effect diminishes one year later \citep{Zhong2020}. However, \cite{NZIER2017} criticize that  the soda tax can be easily circumvented, if the tax is implemented very locally, such as in a particular city. For this reason, \cite{NZIER2017} caution against over-interpreting the results of these studies.

There is little literature that relates to France and Hungary which are analysed in this paper. The soda tax literature based on France reports a pass-through  of the soda tax to the consumer between 39\% \citep{Etile} and 100\% (\cite{Capacci2019} and  \cite{Berardi2016}). \cite{Capacci2019} analyse the number of purchased sodas at the household level in France compared to close regions in Italy. They apply a differences-in-differences approach and find a very small but imprecisely estimated decrease in soda consumption and explain their result with the low soda tax level. \cite{Biro2016} evaluates the effect of the unhealthy product tax in Hungary but sodas are excluded from the analysis due to data restrictions. \cite{Kurz2021} examine both the soda tax in France and the PHPT in Hungary  using data on purchased sodas and apply a synthetic control group analysis. Their findings suggest a slight decline in soda consumption in France and a decrease in Hungary in the short run that fades away two years later. However, these results are imprecisely estimated due to the small number of observations.

The second strand of literature discusses the effect of the soda tax on children's and adolescents' outcomes in other countries than France and Hungary. The evidence is mixed in these studies: \cite{FLETCHER2010} finds  a small decline in soda consumption of children and adolescents due to the implementation of a soda tax across all U.S. states, which had an average soda tax between 1.5\% and 2.3\%.   Whereas \cite{Sturm2010} ascertain that a low soda tax ($\leq$ 4\%) has neither an effect on soda consumption nor on obesity rates of children in the U.S. states. In Philadelphia, a tax of converted  49 Eurocents per litre\footnote{converted into litre and Euros (at the rate on 01.01.2017)} does not affect children's soda consumption \citep{CAWLEY2019}. However, subgroups like overweighted children, children from families with low incomes \citep{Sturm2010} or children with a high soda consumption prior to the reform \citep{CAWLEY2019}  are more likely to react to a soda tax. Regarding the effect of the soda tax on  BMI, \cite{POWELL2009} find no significant change among adolescents in different U.S. states.   In Peru, a price war among the manufactures lead to a reduction of soda prices at the end of the nineties, which increased the obesity rate among children \citep{Ritter}.

To the best of my knowledge, this is the first paper analysing the impact of two different soda taxes  on soda consumption and health of school-aged children.  A cross-country comparison is difficult to draw  because  countries collect their data by themselves with different questions and methods \citep{JOU2012}. Using the health behaviour in school-aged children (HBSC) dataset enables me to draw a cross-country comparison.

The remainder of this paper is organized as follows: In the next section, I provide information about the implementation of the soda tax in France and Hungary. Thereafter, I present the data source, descriptive statistics, and define the subsample. Then, I discuss the empirical strategy and show the results. Finally, I conclude the empirical analysis.

\section{Institutional background: Soda taxes}
\label{s:background}

Soda taxes represent a policy tool to combat sugar intake on a country and local level.  Several U.S. cities, as well as various countries, have implemented a tax on sugar-sweetened sodas  in recent years \citep{Allcott2019}. The first two countries which imposed a soda tax in Europe were  Finland in 1940 followed by Norway  in 1981 \citep{TaxF}.

Thirty years after Norway, Hungary levied a "Public Health Product Tax" (PHPT) on salted snacks, condiments, flavoured alcohol, fruit jams, confectionery, energy drinks, and also on sugar-sweetened beverages \citep{Ecorys2014}. Every product category reveals a different tax level, even among the sugar-containing beverages exist differences: Syrups are taxed by 200 Hungarian Forint (HUF)\footnote{equals 73 Eurocents on 01.09.2011.} per litre, whereas other sugar-sweetened sodas are taxed by 7 HUF\footnote{equals 2.2 Eurocents on 01.01.2012}  per litre. Additionally, the original tax level for sodas amounted to 5 HUF in 2011 and was increased to 7 HUF in 2012 \citep{Biro2016}. This soda tax only affects sodas exceeding a sugar content of 8 grams per 100 millilitres, sodas with less sugar are not taxed. To make this threshold more apparent, an original Coca-Cola contains more than 10 grams of sugar per 100 millilitres.\footnote{https://www.coca-cola.co.uk/our-business/faqs/how-much-sugar-is-in-coca-cola, last retrieved on 07.11.2021.} The reason for the implementation of this  tax was a public health crisis \citep{WHO2015}. Non-communicable diseases have been a widespread cause of premature death, whereby the main risk factor was  unhealthy diet \citep{UniversityW}. One year prior to the implementation of the health policy, the share of overweight adults\footnote{Persons 18 years and older with a body mass index equal or bigger than 25.} amounted to 61.7\% \citep{OverweightbyCountry} which exceeded the European average of 58.7\% \citep{OverweightbyRegion}. Likewise, the share of obese adults was higher (25.3\%) \citep{ObesitybyCountry} than the European average (22.3\%) \citep{ObesitybyRegion}.

 According to the law, the consumer bears the tax burden, by paying retail price including the soda tax \citep{Ecorys2014}. To answer the question whether the soda tax was passed through to the consumer, I tracked the development of the annual average price of two litres of Coca-Cola. Table~\ref{fig:price_dev} reports a price increase of Coca-Cola  from the soda tax implementation in 2011 onwards. With this data at hand, I calculated the yearly price increase and the price index based on the previous year. The price of two litres of Coca-Cola rose from the pre-treatment year 2010 to the implementation year 2011 by 21 HUF. To see whether the price increase was driven by inflation or the soda tax, I compared the Coca-Cola price index with the consumer price index (CPI) of non-alcoholic beverages. The price of Coca-Cola increased more than the price of non-alcoholic beverages in general, which points to the pass-through of the soda tax. In 2012, the tax was slightly raised from 5 HUF to 7 HUF per litre and comparing the two different price indexes shows an even greater difference than in the year before. The price indexes started to converge in the year 2013 and  approximated each other in 2014. 

\begin{table}[!ht]
  \centering
  \caption{Price development of Coca-Cola and consumer price index (CPI) of non-alcoholic beverages}\label{fig:price_dev}
  \label{Price development of Coca-Cola and CPI of non-alcoholic beverages}
  \begin{tabular}{l |l l l l l l}
 & 2010 & 2011 & 2012 & 2013 & 2014\\
 \toprule
Coca-Cola average price (2l) &295 HUF & 316 HUF & 351 HUF & 366 HUF&  373 HUF\\
Coca-Cola  price increase (2l)  &0 HUF & 21 HUF &  35 HUF & 15  HUF & 7 HUF\\
Coca-Cola price index    &100  &107.1 & 111.1 & 104.3 &102.0  \\
  \midrule
CPI of non-alcoholic beverages  &100.1 & 101.9&105.9 &101.9 &100.5 \\
\hline
\end{tabular}

\begin{singlespace}
{\footnotesize Source: Data concerning the  Coca Cola annual average price and the  CPI of non-alcoholic beverages stems from the Hungarian Central Statistical Office. The Coca-Cola price increase and index is self-calculated.}
\end{singlespace}
\end{table}

A beverage tax was implemented in France in January 2012 and it was originally intended to apply to sugar-sweetened sodas only. However, this was, according to \cite{Ecorys2014}, not possible, because the customs codification classifies sodas with added sugar and sweetener into the same category. Whereas \cite{BODO2019} reports that sodas with artificial sweeteners were included in the tax to generate higher tax revenues for the farm sector.  Consequently, both kinds of sodas are taxed, independently of their quantity of sugar or sweetener.
The tax increased over time from 7.16 Eurocents per litre in 2012, to 7.31 Eurocents in 2013, and reached 7.45 Eurocents in 2014 \citep{Ecorys2014}. The aim of the soda tax is twofold: Firstly, it is designed to collect additional revenue for the health  \citep{Ecorys2014} and  farm sector \cite{BODO2019}. Secondly, the soda tax is supposed to reduce the obesity rate among French citizens \citep{Ecorys2014}. One year prior to the implementation of the health policy, France revealed a share of overweight adults exceeding the European average by almost 2 percentage points (\cite{OverweightbyCountry} and \cite{OverweightbyRegion}). Every fifth adult had a BMI equal to or over 30 which indicates obesity among the WHO definition, yet the share of obese adults in France is lower than the European average (\cite{ObesitybyCountry} and \cite{ObesitybyRegion}). Moreover, the National Nutrition and Health Programme 2011 formulated the goal to decrease the share of children drinking more than half a glass of soda a day by at least a quarter in the following five years \citep{Minist2011}. The discussion about the implementation of the soda tax lasted from 2005 to 2011. In August 2011 the tax was decided unexpectedly, the implementation followed five months later \citep{BODO2019}. The pass-through  of the soda tax to the consumer is reported between 39\% \citep{Etile} and 100\%  (\cite{Capacci2019} and  \cite{Berardi2016}).

Table~\ref{fig:comp_taxes} highlights the main differences in the taxes  in  Hungary and France. Hungary's Public Health Product Tax includes a bunch of unhealthy products and not only sodas. However, exclusively sugar-sweetened sodas exceeding the threshold of 8 g sugar per 100 millilitres are taxed, whereas sodas with less sugar or artificial sweetener are exempt from the tax. France taxes sugar-sweetened soft drinks (regardless of sugar content) and soft drinks with artificial sweeteners. The tax level is different among the countries, yet both countries raised the tax after implementation.
\begin{table}[!ht]
  \centering
  \caption{Comparison of taxes in Hungary and France}\label{fig:comp_taxes}
  \begin{tabular}{l |l l}
 & Hungary & France \\
 \toprule
Implementation year & 2011 & 2012\\
\\
Tax  & Unhealthy product tax & Beverage tax \\
    \\
Products & Sodas & Sodas \\
 & Syrups &  \\
 & Energy drinks &  \\
  & Confectionery &  \\
    & Salted snacks &  \\
  & Condiments &  \\
   & Flavoured alcohol &  \\
    & Fruit jams &  \\
    \\
Kind of sodas & Sugar-sweetened & Sugar-sweetened \\
& & or artificial sweetener\\
\\
Threshold of sugar    & $>$ 8 grams per 100 millilitres  & None \\
\\
Tax level in implementation year &  5 HUF ($\sim$ 1.8 Eurocents)   & 7.16 Eurocents   \\
\\
Tax level in evaluation year (2014) &    7 HUF ($\sim$ 2.4 Eurocents)   & 7.45 Eurocents   \\
\\
Tax level increase & 40\% & 4\%\\

\hline
\end{tabular}
\begin{singlespace}
{\footnotesize Source: The information concerning Hungary and France stems from  \cite{Ecorys2014}, the table was created by myself.}
\end{singlespace}
\end{table}

\section{Hypothesis Development}\label{s:data}
Soda taxes aim to raise the price  and to decrease the consumption of sugar-sweetened beverages \citep{NZIER2017}. In theory, manufacturers collect the retail price including the tax from the customer and transfer it to the state. In practice, the manufacturers might decide to reduce their margin and to  pass only part of the tax to the customer. Empirical evidence ranges from a partly pass-through of 39\% \citep{Etile} to 100\%  (\cite{Capacci2019} and  \cite{Berardi2016}) in France.  Table~\ref{fig:price_dev} reveals a price increase of Coca-Cola greater than the inflation after the implementation of the soda tax in Hungary. Consequently, the soda prices increased for the consumer in both countries. Figure~\ref{fig:soda_price_devel} illustrates the increase in the soda price due to the positive pass-through of the tax.\\

\begin{figure}[!htbp]
\centering
 \fbox{\parbox[b][.5\baselineskip]{0.09\linewidth}{Soda tax}}
+
\fbox{\parbox[b][.5\baselineskip]{0.2\linewidth}{Pass-through $>$ 0\%}}
$\Rightarrow$
\fbox{\parbox[b][.5\baselineskip]{0.13\linewidth}{Soda price $\uparrow$}}
\caption{Soda price development}\label{fig:soda_price_devel}
\end{figure}

 Whether the increased price leads to a decrease in demand depends on the price elasticity of  sodas \citep{NZIER2017}. Sodas are goods with a rather high price elasticity because they do not belong to staple food. Hence, I expect a reduction of soda consumption as shown in Figure~\ref{fig:red_soda_con}).  \\

\begin{figure}[!htbp]
\centering
\fbox{\parbox[b][.5\baselineskip]{0.13\linewidth}{Soda price $\uparrow$}} 
+
\fbox{\parbox[b][.5\baselineskip]{0.2\linewidth}{High price elasticity}}
$\Rightarrow$
\fbox{\parbox[b][.5\baselineskip]{0.21\linewidth}{Soda consumption $\downarrow$}}
\caption{Reduction of soda consumption}\label{fig:red_soda_con}
\end{figure}

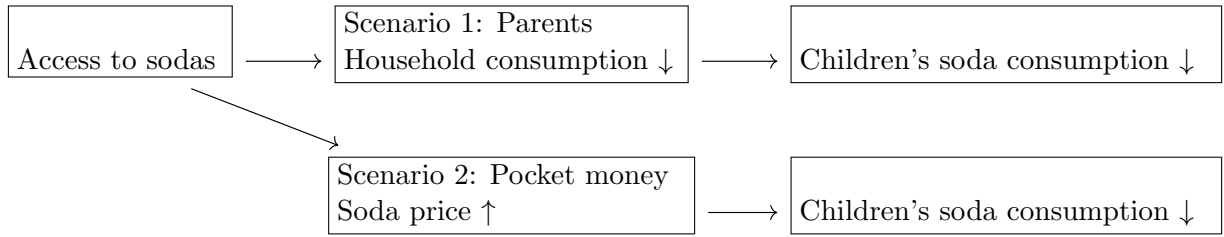
\begin{figure}[!htbp]
\centering
\begin{tikzcd}[ampersand replacement=\&]
\fbox{\parbox[b][1.5\baselineskip]{0.16\linewidth}{Access to sodas}} \arrow[rd] \arrow[r]\& \fbox{\parbox[b][1.5\baselineskip]{0.26\linewidth}{Scenario 1: Parents\\ Household consumption $\downarrow$}} \arrow[r]    \& \fbox{\parbox[b][1.5\baselineskip]{0.32\linewidth}{Children's soda consumption $\downarrow$}}\\
\& \fbox{\parbox[b][1.5\baselineskip]{0.27\linewidth}{Scenario 2: Pocket money\\ Soda price $\uparrow$ }}\arrow[r]  \& \fbox{\parbox[b][1.5\baselineskip]{0.32\linewidth}{Children's soda consumption $\downarrow$}}\\
\end{tikzcd}
\caption{Children's access to sodas}\label{fig:access_to_sodas}
\end{figure}

In a next step, this hypothesis is adjusted to the subgroup of school-aged children which is presented in Figure~\ref{fig:access_to_sodas}. The crucial question is: How do children get mainly access to sugar-sweetened sodas? Either parents provide these beverages or children spend (part of) their pocket money on sodas. In the first case, children's soda consumption should also decrease when household consumption decreases due to the soda price increase. In the second case, children are even more affected by a price increase, because of their limited budget. Hence, I expect a decrease in soda consumption among children.\\

The second hypothesis addresses the effect of the soda tax on children's BMI and depends on the substitution behaviour of the children: If  children consume less sugar due to a lower soda intake and this amount of sugar is not substituted by other sugar-containing products, the average BMI of children might decline. For example, a substitution of sugar-sweetened drinks by sugar-free drinks pointed to a decrease in weight among children \citep{Ruyter2012}. 
However, if children substitute the saved sugar  with other sugary products, the soda tax does not affect the BMI. Since \cite{POWELL2009} find no significant effect of a soda tax on children's BMI in the U.S., I expect at least not an increase in BMI due to the soda tax.

\section{Data}\label{s:data}

To undertake a cross-country comparison, I use data about the health behaviour in school-aged children (HBSC). This survey is conducted in cooperation with the World Health Organization (WHO) Europe on a quadrennial basis since 2001.  The advantage of this survey is the use of the same questionnaire in each country. One question captures the frequency of sugar-sweetened soda consumption on a 7-point scale from "never" to "more than once daily". Another section addresses body measures like children's height and weight. Based on this information, the body mass index (BMI) is calculated and reveals whether a child is overweight. The random sampling is made class-wise which implies a repeated cross-sectional design.

Regarding the sample restriction, I excluded pupils with missing observations in the dependent variable like soda consumption and body mass index. In the survey year 2006  is no information about the BMI available, so I calculate it by the following formula $\text{BMI} = \text{bodyweight}/\text{(body height in m)}^2$.\footnote{https://projekte.uni-hohenheim.de/wwwin140/info/interaktives/bmi.htm} Furthermore, I have excluded children with missing information about their age, sex, TV consumption on weekdays, and having their own bedroom. Moreover, observations are excluded from the sample that do not inform about the number of computers in the household, ownership of a car, family wealth, and whether the mother or the father lives in the main home.

An appropriate control group is similar to the treatment group but without the implemented policy. This approach enables us to attribute the change in the outcome to the implemented policy in the treatment group \citep{Taillie2017}. Therefore, I use a neighbouring country, without a soda tax in force, as a control  group. As Table~\ref{fig:Treatment and control groups} shows, Croatia constitutes the control group for Hungary and Spain for France. Hungary has implemented the tax in 2011 and France followed one year later, yet the data is available every four years. Therefore, the survey year 2010 constitutes the pre-treatment period and 2014 the post-treatment period. In the treatment as well as in the control group are 11 to 15 years old surveyed pupils. \\

\begin{table}[!ht]
  \centering
  \caption{Treatment and control groups}
  \label{fig:Treatment and control groups}
  \begin{tabular}{r cc cc}
    \multicolumn{5}{c}{Panel A: Hungary and Croatia} \\
    \toprule
    \multicolumn{1}{c}{} & \multicolumn{2}{c}{Treatment}& \multicolumn{2}{c}{Control}\\
    &\multicolumn{1}{c}{Country} & \multicolumn{1}{c}{Survey year}& \multicolumn{1}{c}{Country} & \multicolumn{1}{c}{Survey year}
     \\
    \midrule
    \multicolumn{1}{l}{Pre-treatment} & \multicolumn{1}{c}{Hungary} & \multicolumn{1}{c}{2010} &\multicolumn{1}{c}{Croatia} & \multicolumn{1}{c}{2010}\\
    \multicolumn{1}{l}{Post-treatment} & \multicolumn{1}{c}{Hungary} &  \multicolumn{1}{c}{2014} & \multicolumn{1}{c}{Croatia} & \multicolumn{1}{c}{2014}\\
    \bottomrule
    \\
      \multicolumn{5}{c}{Panel B: France and Spain} \\
    \toprule
    \multicolumn{1}{c}{} & \multicolumn{2}{c}{Treatment}& \multicolumn{2}{c}{Control}\\
    &\multicolumn{1}{c}{Country} & \multicolumn{1}{c}{Survey year}& \multicolumn{1}{c}{Country} & \multicolumn{1}{c}{Survey year}
    \\
    \midrule
     \multicolumn{1}{l}{Pre-treatment} & \multicolumn{1}{c}{France} & \multicolumn{1}{c}{2010} &\multicolumn{1}{c}{Spain} & \multicolumn{1}{c}{2010}\\
    \multicolumn{1}{l}{Post-treatment} & \multicolumn{1}{c}{France} &  \multicolumn{1}{c}{2014} & \multicolumn{1}{c}{Spain} & \multicolumn{1}{c}{2014}\\
    \bottomrule
    \\
  \end{tabular}
\end{table}

 Table~\ref{fig:Descriptive statistics: France (treated) and Spain (non-treated)} reports the descriptive statistics for the treated children living in France and the untreated children living in Spain separately. For either group, the mean and the standard deviation (std.dev) of the variables are provided. The last two columns contain the mean differences across groups as well as the p-values. In France, the share of boys and girls in the survey is almost balanced and children are on average 13.5 years old. The children watch on average 2 hours TV on a weekday and the majority has their own bedroom. Most children live together with their mother and father at home. The family posses on average 1.7 cars and 2.3 computers, and reports well-being between "quite well of" and "average". All these before mentioned control variables are statistically significantly different across children in France and Spain. The lower part of Table ~\ref{fig:Descriptive statistics: France (treated) and Spain (non-treated)} presents the descriptive of the outcome variables. The frequency of consumed sodas is measured as a categorical variable, a value of four corresponds to a soda consumption on two to four days a week. A body mass index (BMI) of 19 is within the normal range.\footnote{https://www.stanfordchildrens.org/en/topic/default?id=determining-body-mass-index-for-teens-90-P01598} The mean differences of the two outcome variables are statistically significant between children in France and Spain. Regarding the sample size, 8,821 children participated in the survey in France and 9,744 children participated in Spain which sums up to 18,565 observations in total.\\

\begin{table}[ht]
\center
\caption{Descriptive statistics: France (treated) and Spain (non-treated)}
\label{fig:Descriptive statistics: France (treated) and Spain (non-treated)}
\begin{center}
{\footnotesize
\begin{tabular}{l|cc|cc|cc}
  \hline\hline
   & \multicolumn{2}{|c|}{Treated} & \multicolumn{2}{c|}{Non-treated} & & \\
  \hline
 & mean & std.dev & mean & std.dev & mean difference & p-value \\ 
\hline
& \multicolumn{4}{c|}{Time} \\
\hline
Year & 2011.97 & 2.00 & 2012.32 & 1.97 & -0.35 & 0.00 \\ 
\hline
& \multicolumn{4}{c|}{Control variables} \\
\hline
\textit{Child characteristics} & & & & & & \\
  Female (Dummy) & 0.51 & 0.50 & 0.52 & 0.50 & -0.02 & 0.04 \\ 
  Age (in years) & 13.54 & 1.65 & 13.68 & 1.61 & -0.14 & 0.00 \\ 
  TV consumption on a weekday (categorical) & 2.09 & 1.77 & 2.00 & 1.60 & 0.09 & 0.00 \\ 
  \hline
\textit{Household characteristics} & & & & & &\\
  Mother living at main home (Dummy) & 0.92 & 0.27 & 0.96 & 0.20 & -0.04 & 0.00 \\ 
    Father living at main home (Dummy) & 0.75 & 0.44 & 0.83 & 0.37 & -0.09 & 0.00 \\ 
   Number of family cars & 1.69 & 0.53 & 1.54 & 0.58 & 0.15 & 0.00 \\ 
   Own bedroom (Dummy) & 0.85 & 0.36 & 0.83 & 0.38 & 0.02 & 0.00 \\ 
   Number of computers per family & 2.30 & 0.80 & 2.23 & 0.83 & 0.07 & 0.00 \\ 
Family well-off (categorical) & 2.28 & 0.85 & 2.93 & 0.50 & -0.65 & 0.00 \\ 
   \hline
& \multicolumn{4}{c|}{Outcome variables} \\
\hline
    Frequency of sodas  (categorical)& 3.97 & 1.87 & 3.85 & 1.76 & 0.12 & 0.00 \\ 
  Body mass index (BMI) & 18.97 & 3.18 & 19.94 & 3.30 & -0.97 & 0.00 \\ 
 Number of observations  & 8,821 &  & 9,744 &  &  &  \\ 
  \hline
\end{tabular}
}
 {\footnotesize Source: Health behaviour of school-aged children (HBSC)\par}
\end{center}
\par

\end{table}

\begin{table}[ht]
\center
\caption{Descriptive statistics: Hungary (treated) and Croatia (non-treated)}
\label{tab:Descriptive statistics: Hungary (treated) and Croatia (non-treated)}
\begin{center}
{\footnotesize
\begin{tabular}{l|cc|cc|cc}
  \hline\hline
   & \multicolumn{2}{|c|}{Treated} & \multicolumn{2}{c|}{Non-treated} & & \\
  \hline
 & mean & std.dev & mean & std.dev & mean difference & p-value \\ 
\hline
& \multicolumn{4}{c|}{Time} \\
\hline
Year & 2011.74 & 1.98 & 2011.64 & 1.97 & 0.09 & 0.00 \\ 
\hline
& \multicolumn{4}{c|}{Control variables} \\
\hline
\textit{Child characteristics} & & & & & & \\
  Female (Dummy) & 0.52 & 0.50 & 0.51 & 0.50 & 0.00 & 0.64 \\ 
  Age & 13.59 & 1.65 & 13.68 & 1.66 & -0.10 & 0.00 \\ 
   TV consumption on a weekday (categorical) & 2.04 & 1.65 & 2.48 & 1.74 & -0.43 & 0.00 \\ 
     \hline
\textit{Household characteristics} & & & & & &\\
  Mother living at main home (Dummy) & 0.95 & 0.23 & 0.98 & 0.13 & -0.04 & 0.00 \\ 
  Father living at main home (Dummy) & 0.74 & 0.44 & 0.94 & 0.24 & -0.20 & 0.00 \\ 
  Number of family cars & 1.04 & 0.71 & 1.34 & 0.59 & -0.30 & 0.00 \\ 
  Own bedroom (Dummy) & 0.73 & 0.44 & 0.67 & 0.47 & 0.07 & 0.00 \\ 
Number of computers per family & 1.83 & 0.87 & 1.73 & 0.85 & 0.10 & 0.00 \\ 
Family well-off (categorical) & 2.40 & 0.83 & 2.15 & 0.91 & 0.25 & 0.00 \\ 
  \hline
& \multicolumn{4}{c|}{Control variables} \\
\hline
    Frequency of sodas  (categorical) & 4.03 & 1.98 & 3.94 & 1.84 & 0.09 & 0.00 \\ 
  Body mass index (BMI) & 19.54 & 3.49 & 19.90 & 3.24 & -0.36 & 0.00 \\ 
  Number of observations & 7,544 &  & 9,919 &  &  &  \\ 
  \hline
\end{tabular}
}
 {\footnotesize Source: Health behaviour of school-aged children (HBSC)\par}
\end{center}
\par

\end{table}

Table~\ref{tab:Descriptive statistics: Hungary (treated) and Croatia (non-treated)} reports the descriptive statistics for the second country-pair. Children living in Hungary belong to the treatment group and children living in Croatia form the control group. TV consumption is slightly higher among the children in Croatia and it is more likely that the father lives in the main household in Croatia. Families in Croatia have a higher probability to have more than one car in comparison to families in Hungary, whereas children in Hungary are more likely to have their own bedroom. The number of computers at home is slightly higher in Hungary than in Croatia, whereas the families in Croatia score higher in being well off. Children in Hungary consume slightly more frequently sodas than children in Croatia. The BMI is slightly higher in Croatia than in Hungary. There are 7,544 treated children in Hungary and 9,919 non-treated children in Croatia, so 17,463 children in total.

\section{Econometric approach}\label{s:metrics}

In this chapter, I discuss the Difference-in-Differences (DiD) strategy for identifying the Average Treatment Effect on the Treated (ATET) (see e.g. \cite{Lechner10}), i.e. among children living in Hungary or France  in 2014. The potential outcome Y (e.g. frequency of consumed sodas) depends on the time period t $\in$ \{0, 1\} and the potential treatment state d  	$\in$ \{0, 1\} . The notation  $Y_{t}^{d}$  indicates the potential outcome in the potential treatment state d and in time period t. For example, the potential outcome of the treatment group (d = 1) in the pre-treatment period (t = 0) is represented by  $Y^{1}_{0}$. This notation facilitates to state the identifying assumptions of the DiD framework, see \cite{Lechner10}:


The first assumption, formulated in equation \ref{exon}, implies the exogeneity of the covariates (X). This assumption would be violated if the soda tax affects the characteristics of the children or the household. Time-independent covariates,  like gender, cannot be affected by the soda tax because they are constant over time. Time-dependent variables may be affected by the treatment, especially if these variables are measured after the implementation of the soda tax. Since I use  repeated cross-sections, the covariates are measured in 2014, whereas the soda tax  is in force since 2011 or 2012 respectively. However, it is rather unlikely that the soda tax affects, for example, children's TV consumption or whether the mother lives at the main home or not.

\begin{equation}
 \begin{array}{l}
X^{1}=X^{0}=X; \; \forall x   	\in  \chi.\label{exon}
\end{array}
\end{equation}

The main identifying assumption in the context of DiD is the common trend assumption, formally stated in equation \ref{com}. Intuitively speaking, the soda consumption and the BMI of children living in Hungary  and Croatia, would follow the same time trend in the absence of the soda tax.\footnote{This  assumption must hold  for France and Spain too.} For this reason, I need to control for child and household covariates that would lead to different time trends. For example, boys and children from low-income families consume more sugar-sweetened sodas than girls and children from more privileged families \citep{Schroeder2021}, hence the time trend differs between these groups. I provide a placebo test conditional on covariates using unaffected periods  in Table \ref{tbl:unaf_periods_soda} in Section~\ref{s:results_soda} to support this assumption.

\begin{equation}
 \begin{array}{l}
E[Y^{1}_{0}|X=x, D=1] - E[Y^{0}_{0}|X=x, D=1]= \\
E[Y^{1}_{0}|X=x, D=0] - E[Y^{0}_{0}|X=x, D=0]=\\
E[Y^{1}_{0}|X=x] - E[Y^{0}_{0}|X=x]; \; \forall x   	\in  \chi.\label{com}
\end{array}
\end{equation}

A further assumption rules out an anticipatory effect ($\theta$) of the policy  in the pre-treatment period (t = 0) as formulated in equation \ref{antic}. Accordingly, children in the treated countries Hungary and France must not anticipate the effect of the soda tax in 2010 and their soda consumption must not change prior to the implementation of the tax. Since the tax was discussed from 2005 to 2011 in the parliament in France, it might have raised the awareness of unhealthy beverages among the French children. However, the decision to pass this law was unexpected and the implementation time of five months was rather short  \citep{BODO2019}. In Hungary, the law was passed one and a half months before it came into force \citep{Ecorys2014}, which represents also a short period for anticipatory behaviour.

\begin{equation}
 \begin{array}{l}
\theta_{0}(x)= 0; \; \forall x   	\in  \chi.\label{antic}
\end{array}
\end{equation}

The last assumption is known as the common support assumption and is formulated in equation \ref{cosu}. It ensures that for each child in Hungary in the year 2014, another child exists with the same characteristics in the following three groups: i) Hungary in 2010, ii) Croatia in 2010, and iii) Croatia in 2014.\footnote{This assumption holds for  France in 2014 too. In this case, the three groups are i) France in 2010, ii) Spain in 2010, and iii) Spain in 2014.} 
Under assumptions \ref{exon} - \ref{cosu}, the ATET is identified.

\begin{equation}
 \begin{array}{l}
P[TD=1|X=x, (T,D)	\in {(t,d), (1,1)}]< 1; \\ \; \forall (t, d)   	\in  \{(0,1), (0,0), (1,0)\}; \;\forall x   	\in  \chi.\label{cosu}
\end{array}
\end{equation}

A standard DiD approach models a linear relationship between the policy and the outcome, in this case, the outcome variable is continuous.  The variable "Frequency of sodas" is measured as a categorical variable in the HBSC dataset. Therefore, this variable is a limited dependent variable, implying a non-linear relationship between the policy and the outcome. However, considering the non-linearity may lead to the violation of the identifying assumption of the DiD, the common trend assumption \citep{Lechner10}. To deal with this issue, I use a semi-parametric approach to model the relationship in a more flexible way than a parametric approach. 

Equation \ref{idi} describes the identification of the semi-parametric ATET based on inverse probability weighting \citep{Huber2019}. The outcome variable Y is multiplied by an inverse probability weight, where $\Pi$ gives the share of treated observations in the post-treatment period and  $\rho_{d,t}(X)$ is the probability of being in the treatment state d and in the time period t, conditional on  covariates X. This propensity score is estimated by probit.

\begin{equation}
 \begin{array}{l}
E\left [ \left \{ \frac{DT}{\Pi}- \frac{D(1-T)\rho _{1,1}(X)}{\rho _{1,0}(X)\Pi }- (\frac{(1-D)T\rho _{1,1}(X)}{\rho _{0,1}(X)\Pi })-\frac{(1-D)(1-T)\rho _{1,1}(X)}{\rho _{0,0}(X)\Pi }) \right \}Y \right ],\\
where\  \Pi = Pr(D=1, T=1),\ \rho _{d,t}(X)=Pr(D=d, T=t|X).\label{idi}
\end{array}
\end{equation}

To ensure that the common trend assumption holds, I include the following covariates (X) in the estimation: On the individual level, I control for age and sex of the child, because older children reveal a different soda consumption than younger children and boys differ in their consumption behaviour from girls \citep{Vereecken2005}.  Since TV consumption was associated with soda consumption (see for example \cite{ANDREYEVA2011} and \cite{GRIMM2004}), I control for  television consumption on a weekday. On the household level, I take into account several characteristics: Firstly, I control for the household structure, in particular, whether the mother or the father lives in the same household as the child. Secondly, I control for the wealth of the family, because it is associated with different soda consumption levels \citep{DREWNOWSKI2019}. I use the following proxies for family's wealth: Ownership of a family car, number of computers in the household, well-off of the family, a dummy indicating whether the child has his/her own bedroom. Furthermore, soda consumption increases with wealth in Eastern European countries, whereas it decreases in Western European countries \citep{Vereecken2005}. Country-specific characteristics, like the growth of the Gross Domestic Product (GDP), may affect the soda consumption of its inhabitants and thus bias the results. Controlling for country-specific covariates could serve as a solution for this problem, yet this is  not possible because of the multi-collinearity with the treatment. Therefore, I inspect the GDP growth of each country pair in Section~\ref{s:results_soda}.

For the estimation, I use the didweight command of the causalweight package in R, with the default number of bootstrap replications of 1,999 to calculate the standard errors, and the default trimming rule of 0.05 to drop observations with an extreme propensity score from the sample. Furthermore, I use no clusters in the analysis. Clustered standard errors would help to get consistent standard errors given the number of clusters is large enough. However, in this setting there is only one country in the treatment group. I do not use clustered standard errors on the country level because whenever bootstrap does not draw the one treated country, the observation is dropped. In the Appendix, school-year specific clustering is used as a robustness check. Since the didweight command is designed for one pre- and one post-treatment-period, I use the survey years 2010 and 2014 in the estimation. Several pre-treatment years are available to test the parallel trend assumption. I use the survey years 2006 and 2010 and run the estimation with a fake treatment in the latter.

\section{Results}\label{s:results_soda}
This chapter provides the estimated results as well as the sensitivity analysis. Table~\ref{tbl: emp_results} presents the effects of the implemented tax, the standard errors  are estimated by bootstrap, the p-values are obtained from t-tests. Panel A reports the effect of the policy package in Hungary, whereas Croatia constitutes the control group.  The findings point at the first glance to a counter-intuitive positive and significant (p $<$ 0.01) effect of the tax on consumption behaviour among school-aged children.\footnote{The effect size is not directly interpretable because soda consumption is measured as a categorical variable.} Since a range of products are taxed, the substitution of sugar-sweetened products might drive this result. For example, a survey among adults  who changed their nutritional behaviour due to the PHPT suggested that 52\% substituted energy drinks with sodas \citep{WHO2016}, which might be driven by the higher tax on energy drinks.\footnote{250 HUF/l if taurin $>$ 100 mg per 100 ml or 40 HUF/l if no taurin but methylxanthine $>$ 15 mg per 100 ml see \cite{BIRO2015}.} However,  another driver of this result might be the relative stronger GDP growth in Hungary compared to Croatia (see Figure~\ref{GDPgrowthHUCR} in the Appendix), i.e. children in Hungary could spend more money for sodas.\footnote{Due to the multicollinearity with the treatment, I cannot control for the GDP growth.}
The second outcome, children's BMI is not affected by the tax in Hungary. Panel B in Table~\ref{tbl: emp_results} reports the effects of the soda tax on the frequency of sugar-sweetened soda consumption and the BMI for French children (treated) and Spanish children (untreated). The effects have the expected negative sign but are insignificant. This result is consistent with analyses of large quantities of soft drinks purchased at the aggregate level such as households \citep{Capacci2019} and industry \citep{Kurz2021}, which find only a small and imprecisely estimated decrease in soft drinks purchased. But Table~\ref{fig: parallel GDP growth in F and Sp} shows larger GDP growth in Spain than in France over time which might affect the result.

\begin{table}[!ht]
  \centering
  \caption{Empirical results}
  \label{tbl: emp_results}
  \begin{tabular}{p{3cm} p{1.5cm} p{2.5cm} p{1.5cm}p{4cm}}
    \multicolumn{5}{c}{Panel A: Hungary and Croatia} \\
    \toprule
     & Effect & Standard error & P-value& Number of observations \\
    \midrule
    \multicolumn{1}{l}{Frequency of sodas} & \multicolumn{1}{c}{0.35} & \multicolumn{1}{c}{0.07} & \multicolumn{1}{c}{0.00} & \multicolumn{1}{c}{18,712}\\
    \multicolumn{1}{l}{Body Mass Index (BMI)} & \multicolumn{1}{c}{0.12} & \multicolumn{1}{c}{0.13} & \multicolumn{1}{c}{0.36}& \multicolumn{1}{c}{17,553}\\
    \bottomrule
     \\
    \multicolumn{5}{c}{Panel B: France and Spain} \\
     \toprule
     & Effect & Standard error & P-value& Number of observations \\
    \midrule
    \multicolumn{1}{l}{Frequency of sodas} & \multicolumn{1}{c}{-0.08} & \multicolumn{1}{c}{ 0.08 } & \multicolumn{1}{c}{0.31} & \multicolumn{1}{c}{ 20,951} \\
    \multicolumn{1}{l}{Body Mass Index (BMI)} & \multicolumn{1}{c}{-0.07} & \multicolumn{1}{c}{0.15} & \multicolumn{1}{c}{0.66} & \multicolumn{1}{c}{18,723} \\
    \bottomrule
    \\
    \par
      \end{tabular}\\
{\footnotesize Note: Standard errors are estimated by bootstrap.}

\end{table}

A downside of non-clustered standard errors in a DiD setting is the possibility of underestimating the standard errors \citep{Duflo2004}. To check whether the empirical results in Table~\ref{tbl: emp_results} are robust, I have re-estimated the results with clustered standard errors. The most conservative approach would be to cluster standard errors  on an aggregate level. However, it is not possible to cluster on the country level  in this setting, because of the few numbers of treated countries. Whenever cluster bootstrap does not draw the treatment group, the procedure does not work. Hence I cluster on the next lowest level which has variation in the data: the school-year level. Table~\ref{tbl:school-year specific clusters} in the Appendix shows that the clustered standard errors are almost equal to the robust standard errors in Table~\ref{tbl: emp_results}.

The identifying assumption of the DiD approach is the parallel trend assumption, implying that, conditional on the covariates, the treatment and control group follow the same time trend in the absence of the treatment. This assumption is not testable, yet it is possible to conduct a placebo test to support this assumption: I use the two pre-treatment periods 2006 and 2010 and pretend that in the latter a 'fake treatment' was implemented. Table~\ref{tbl:unaf_periods_soda} reports large p-values in both panels, which supports the parallel trend assumption.

\begin{table}[!ht]
  \centering
  \caption{Unaffected periods}
  \label{tbl:unaf_periods_soda}
  \begin{tabular}{p{3cm} p{1.5cm} p{2.5cm} p{1.5cm}p{4cm}}
    \multicolumn{5}{c}{Panel A: Hungary and Croatia} \\
    \toprule
    & Effect & Standard error & P-value& Number of observations \\
    \midrule
    \multicolumn{1}{l}{Frequency of sodas} & \multicolumn{1}{c}{0.10} & \multicolumn{1}{c}{0.08} & \multicolumn{1}{c}{ 0.23 } & \multicolumn{1}{c}{19,069}\\
    \multicolumn{1}{l}{Body Mass Index (BMI)} & \multicolumn{1}{c}{-0.06} & \multicolumn{1}{c}{ 0.15} & \multicolumn{1}{c}{0.67}& \multicolumn{1}{c}{ 17,949}\\
    \bottomrule
    \\
    \multicolumn{5}{c}{Panel B: France and Spain} \\
     \toprule
  & Effect & Standard error & P-value& Number of observations \\
    \midrule
    \multicolumn{1}{l}{Frequency of sodas} & \multicolumn{1}{c}{0.09} & \multicolumn{1}{c}{ 0.08 } & \multicolumn{1}{c}{0.26} & \multicolumn{1}{c}{24,919} \\
    \multicolumn{1}{l}{Body Mass Index (BMI)} & \multicolumn{1}{c}{0.01} & \multicolumn{1}{c}{0.15} & \multicolumn{1}{c}{ 0.96} & \multicolumn{1}{c}{ 21,826} \\
    \bottomrule
    \\
    \par
  \end{tabular}\\
  {\footnotesize Note: Standard errors are estimated by bootstrap.}
\end{table}

As a robustness test, I use another neighbouring country  as an alternative control group  in each panel: I substitute Croatia with Slovakia and Spain with Switzerland. Table~\ref{tab:Descriptive statistics: Hungary (treated) and Slovakia (non-treated)} and Table~\ref{tab:Descriptive statistics: France (treated) and Switzerland (non-treated)} in the Appendix report the descriptive statistics of the children and the household for these groups. Table~\ref{tbl:una_periods_other_controlcountry} in the Appendix shows the placebo test of the unaffected periods for Hungary and Slovakia (Panel A) and France as well as Switzerland (Panel B) separately.  It suggests that the parallel trends assumption holds for both Panels, except for the outcome variable BMI in Panel A.   Finally, Table~\ref{tbl:rob_test} reports the results for robustness test.  Even if the control group changes we find a highly statistically significant positive effect on the frequency of consumed sodas in Hungary. In France,  the results on the soda consumption and the BMI have a negative sign,  yet they are insignificant as in the main results in Table~\ref{tbl: emp_results}.  Figure~\ref{fig: parallel GDP growth in H and Slo} and  ~\ref{GDPgrowthFSw} in the Appendix  report  a parallel GDP growth of each country pair prior to the measured effect in 2014. Therefore, GDP growth can be excluded as a driver for the increase in soda consumption in Hungary.

\begin{table}[!ht]
  \centering
  \caption{Robustness test}
  \label{tbl:rob_test}
    \begin{tabular}{p{3cm} p{1.5cm} p{2.5cm} p{1.5cm}p{4cm}}
    \multicolumn{5}{c}{Panel A: Hungary and Slovakia} \\
    \toprule
  & Effect & Standard error & P-value& Number of observations \\
    \midrule
    \multicolumn{1}{l}{Frequency of sodas} & \multicolumn{1}{c}{ 0.48 } & \multicolumn{1}{c}{ 0.09 } & \multicolumn{1}{c}{0.00 } & \multicolumn{1}{c}{ 15,425}\\
    \multicolumn{1}{l}{Body Mass Index (BMI)} & \multicolumn{1}{c}{ -0.13} & \multicolumn{1}{c}{ 0.15  } & \multicolumn{1}{c}{ 0.38 } & \multicolumn{1}{c}{14,059} \\
    
    \bottomrule
    \\
    \multicolumn{5}{c}{Panel B: France and Switzerland} \\
     \toprule
   & Effect & Standard error & P-value& Number of observations \\
    \midrule
    \multicolumn{1}{l}{Frequency of sodas} & \multicolumn{1}{c}{-0.04} & \multicolumn{1}{c}{ 0.07} & \multicolumn{1}{c}{ 0.52  } & \multicolumn{1}{c}{ 22,986} \\
    \multicolumn{1}{l}{Body Mass Index (BMI)} & \multicolumn{1}{c}{ -0.13} & \multicolumn{1}{c}{ 0.11} & \multicolumn{1}{c}{   0.24 } & \multicolumn{1}{c}{20,258} \\
    \bottomrule
    \\
    \par

  \end{tabular}\\
  {\footnotesize Note: Standard errors are estimated by bootstrap.}
\end{table}

\section{Conclusion}\label{s:concl}

This paper examines the effect of two different health policies on sugar-sweetened soda consumption behaviour and body mass index (BMI) of school-aged children in Europe. Hungary has implemented a Public Health Product Tax (PHPT) on several unhealthy products, including sugar-sweetened sodas, in 2011, while France only taxes sodas, containing sugar and artificial sweetener, since 2012. Methodologically, I apply a Difference-in-Differences (DiD) approach to evaluate this natural experiment and use neighbouring countries without such a soda tax as a control group. I analyse the effect in Hungary and France separately, because of the different policy designs. Since the frequency of soda consumption is measured by a categorical scale,  I use a semi-parametric method to estimate the effect in a flexible way. To the best of my knowledge, this is the first paper analysing the impact of two different soda taxes  on the consumption and health of school-aged children.

The results suggest that the PHPT had a statistically significant effect  (p $<$ 0.01) on soda consumption of school-aged children in Hungary, yet the sign is unexpectedly positive. An explanation for this counter-intuitive result might be the substitution behaviour among children, as the price of other unhealthy products, such as energy drinks or syrups, are taxed even higher. In France, the soda consumption of school-aged children is not affected by the soda tax. This result is in line with the analyses of \cite{Capacci2019} and \cite{Kurz2021}, who use soda sales data at a more aggregated level and find a very small, but not robust effect on the quantity of purchased sodas. \cite{Capacci2019} explain this finding  by the very low tax level of 7.16 Eurocents  per litre.

 Moreover, I analyse the effect on children's BMI and find  neither in France nor in Hungary a statistically significant effect. This finding is consistent with  \cite{POWELL2009} who analyses the effect of soda tax on BMI among adolescents. Regarding the sensitivity analysis, I run a placebo test with two unaffected periods. The results suggest an insignificant effect, which supports the parallel trend assumption. The results are robust to an alternative control group.

 Consequently, policy makers should think carefully about the design and the tax rate before implementing a soda tax. The availability of data about children's quantity of soda consumption would help to estimate the effect of the soda tax in a more precise way.



\clearpage
\newpage
\begin{appendices}

\section{Acknowledgment}\label{s:ackn}
HBSC is an international study carried out in collaboration with WHO/EURO. The International Coordinator of the 2005/06 survey was Prof. Candace Currie and the Data Bank Manager was Prof. Oddrun Samdal. 

\newpage

\section{Appendix}\label{s:app}

\renewcommand\thefigure{2.\thesection.\arabic{figure}}    

\begin{figure}[h!]
\centering
\includegraphics[width=0.75\textwidth]{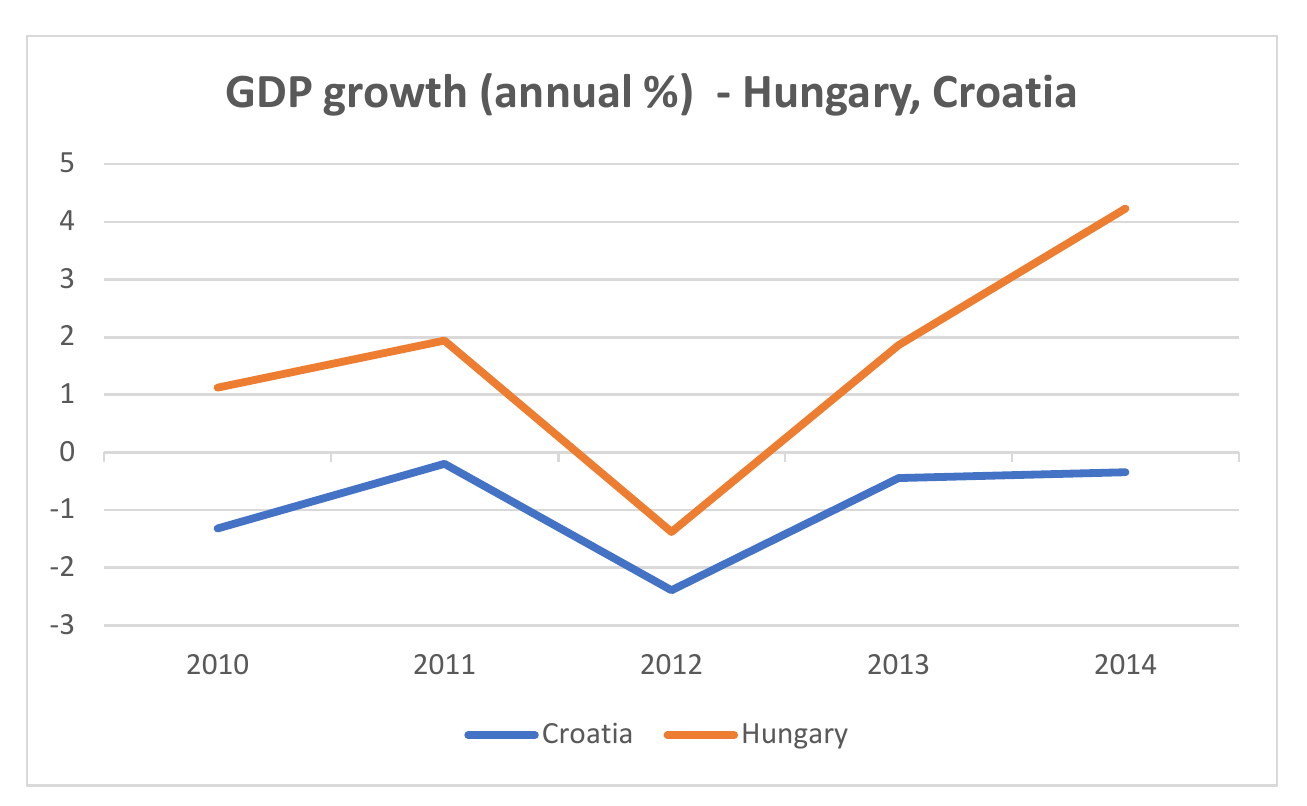}\\
\caption{GDP growth in Hungary and Croatia}
\label{GDPgrowthHUCR}
{\footnotesize Source: The World Bank\footnotemark}
\end{figure}

\footnotetext{\url{https://data.worldbank.org/indicator/NY.GDP.MKTP.KD.ZG?end=2014&locations=HU-HR&start=2010}, last accessed: 10.11.2021.}

\vskip 0.75 cm

\begin{figure}[h!]
\centering
\includegraphics[width=0.75\textwidth]{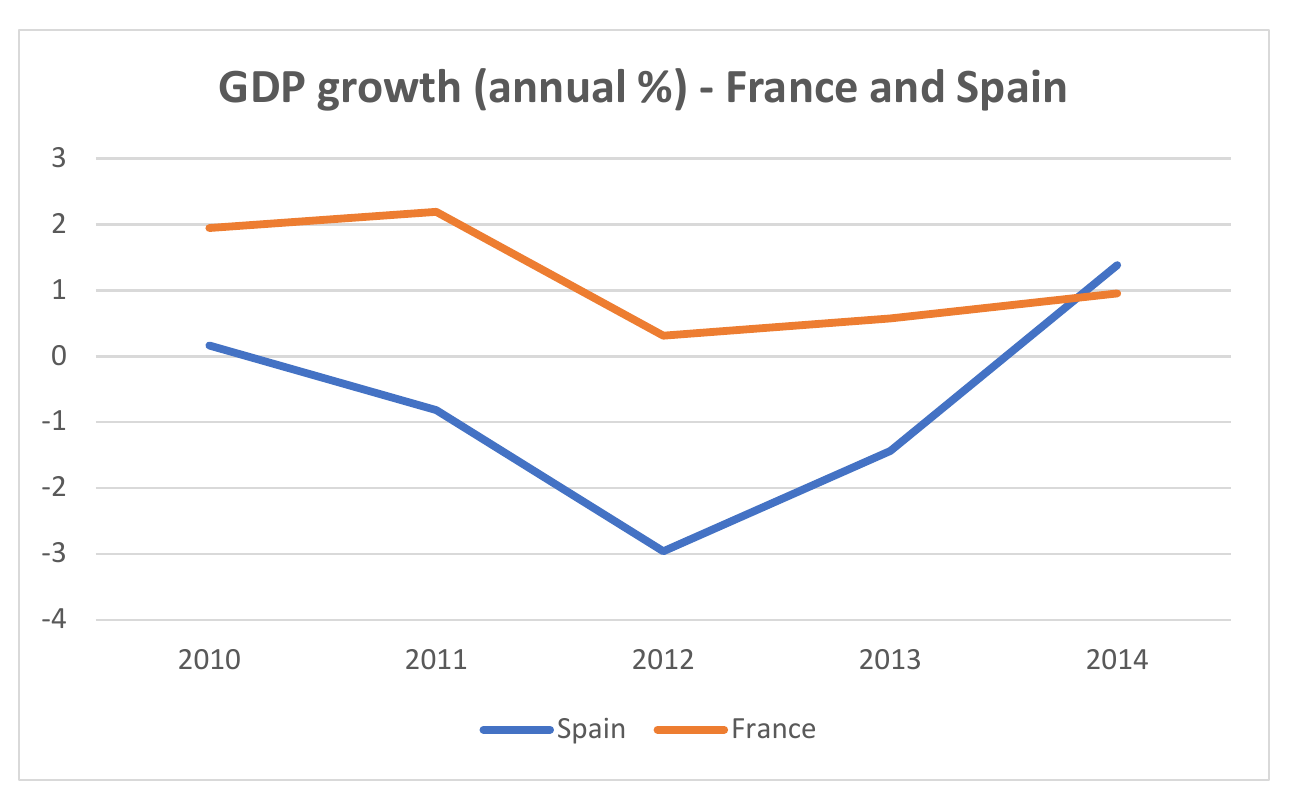}\\
\caption{GDP growth in France and Spain}
\label{fig: parallel GDP growth in F and Sp}
{\footnotesize Source: The World Bank\footnotemark}
\end{figure}

\footnotetext{\url{https://data.worldbank.org/indicator/NY.GDP.MKTP.KD.ZG?end=2014&locations=FR-ES&start=2010}, last accessed: 10.11.2021.}
\newpage
\clearpage

\renewcommand\thetable{2.\thesection.\arabic{table}}    

\begin{table}[!ht]
  \centering
  \caption{Results with school-year specific clusters}
  \label{tbl:school-year specific clusters}
  \begin{tabular}{p{3cm} p{1.5cm} p{2.5cm} p{1.5cm}p{4cm}}
    \multicolumn{5}{c}{Panel A: Hungary and Croatia} \\
    \toprule
    & Effect & Standard error & P-value& Number of observations \\
    \midrule
    \multicolumn{1}{l}{Frequency of sodas} & \multicolumn{1}{c}{0.35} & \multicolumn{1}{c}{0.10} & \multicolumn{1}{c}{0.01} & \multicolumn{1}{c}{18,712}\\
    \multicolumn{1}{l}{Body Mass Index (BMI)} & \multicolumn{1}{c}{0.12} & \multicolumn{1}{c}{0.15} & \multicolumn{1}{c}{0.42}& \multicolumn{1}{c}{17,553}\\
    \bottomrule
     \\
    \multicolumn{5}{c}{Panel B: France and Spain} \\
     \toprule
    & Effect & Standard error & P-value& Number of observations \\
    \midrule
    \multicolumn{1}{l}{Frequency of sodas} & \multicolumn{1}{c}{-0.08} & \multicolumn{1}{c}{ 0.09 } & \multicolumn{1}{c}{0.36} & \multicolumn{1}{c}{ 20,951} \\
    \multicolumn{1}{l}{Body Mass Index (BMI)} & \multicolumn{1}{c}{-0.07} & \multicolumn{1}{c}{0.15} & \multicolumn{1}{c}{0.67} & \multicolumn{1}{c}{18,723} \\
    \bottomrule
    \\
    \par

  \end{tabular}\\
  {\footnotesize Note: Bootstrapped and clustered standard errors.}
\end{table}

\clearpage
\newpage

\begin{table}[ht]
\center
\caption{Descriptive statistics: Hungary (treated) and Slovakia (non-treated)}
\label{tab:Descriptive statistics: Hungary (treated) and Slovakia (non-treated)}
\begin{center}
{\footnotesize
\begin{tabular}{l|cc|cc|cc}
  \hline\hline
   & \multicolumn{2}{|c|}{Treated} & \multicolumn{2}{c|}{Non-treated} & & \\
  \hline
 & mean & std.dev & mean & std.dev & mean difference & p-value \\ 
\hline
& \multicolumn{4}{c|}{Time} \\
\hline
year & 2011.74 & 1.98 & 2011.53 & 1.94 & 0.21 & 0.00 \\ 
\hline
& \multicolumn{4}{c|}{Control variables} \\
\hline
Female & 0.52 & 0.50 & 0.53 & 0.50 & -0.01 & 0.37 \\ 
 Age & 13.59 & 1.65 & 13.98 & 1.35 & -0.39 & 0.00 \\ 
TV consumption on a weekday  & 2.04 & 1.65 & 2.64 & 1.77 & -0.60 & 0.00 \\ 
Mother living at main home (Dummy) & 0.95 & 0.23 & 0.96 & 0.19 & -0.02 & 0.00 \\ 
Father living at main home (Dummy) & 0.74 & 0.44 & 0.88 & 0.33 & -0.14 & 0.00 \\ 
Number of family cars & 1.04 & 0.71 & 1.24 & 0.66 & -0.20 & 0.00 \\ 
Own bedroom (Dummy) & 0.73 & 0.44 & 0.58 & 0.49 & 0.15 & 0.00 \\ 
Number of computers per family & 1.83 & 0.87 & 1.95 & 0.87 & -0.12 & 0.00 \\ 
Family well-off & 2.40 & 0.83 & 2.07 & 0.82 & 0.33 & 0.00 \\ 
\hline
& \multicolumn{4}{c|}{Outcome variables} \\
\hline
    Frequency of sodas & 4.03 & 1.98 & 4.36 & 1.89 & -0.33 & 0.00 \\ 
Body mass index (BMI) & 19.54 & 3.49 & 19.54 & 3.07 & 0.00 & 0.99 \\ 
   & 7,544 &  & 6,419 &  &  &  \\ 
 \hline
 \hline
\end{tabular}
}
 {\footnotesize Source: Health behaviour of school-aged children (HBSC)\par}
\end{center}
\par

\end{table}

\begin{table}[ht]
\center
\caption{Descriptive statistics: France (treated) and Switzerland (non-treated)}
\label{tab:Descriptive statistics: France (treated) and Switzerland (non-treated)}
\begin{center}
{\footnotesize
\begin{tabular}{l|cc|cc|cc}
  \hline\hline
   & \multicolumn{2}{|c|}{Treated} & \multicolumn{2}{c|}{Non-treated} & & \\
  \hline
 & mean & std.dev & mean & std.dev & mean difference & p-value \\ 
\hline
& \multicolumn{4}{c|}{Time} \\
\hline
year & 2011.97 & 2.00 & 2011.93 & 2.00 & 0.04 & 0.15 \\ 
\hline
& \multicolumn{4}{c|}{Control variables} \\
\hline
Female & 0.51 & 0.50 & 0.50 & 0.50 & 0.01 & 0.16 \\ 
Age & 13.54 & 1.65 & 13.58 & 1.57 & -0.04 & 0.08 \\ 
TV consumption on a weekday & 2.09 & 1.77 & 1.50 & 1.35 & 0.59 & 0.00 \\ 
Mother living at main home (Dummy) & 0.92 & 0.27 & 0.97 & 0.18 & -0.05 & 0.00 \\ 
Father living at main home (Dummy)  & 0.75 & 0.44 & 0.81 & 0.39 & -0.06 & 0.00 \\ 
Number of family cars  & 1.69 & 0.53 & 1.46 & 0.60 & 0.24 & 0.00 \\ 
Own bedroom (Dummy)  & 0.85 & 0.36 & 0.87 & 0.33 & -0.03 & 0.00 \\ 
Number of computers per family & 2.30 & 0.80 & 2.41 & 0.76 & -0.11 & 0.00 \\ 
Family well-off & 2.28 & 0.85 & 2.50 & 0.84 & -0.22 & 0.00 \\ 
\hline
& \multicolumn{4}{c|}{Outcome variables} \\
\hline
    Frequency of sodas  & 3.97 & 1.87 & 4.11 & 1.85 & -0.14 & 0.00 \\ 
Body mass index (BMI)  & 18.97 & 3.18 & 19.01 & 3.04 & -0.04 & 0.38 \\ 
   & 8,821 &  & 11,352 &  &  &  \\ 
   \hline
 \hline
\end{tabular}
}
 {\footnotesize Source: Health behaviour of school-aged children (HBSC)\par}
\end{center}
\par

\end{table}

\newpage

\begin{table}[!ht]
  \centering
  \caption{Unaffected periods}
  \label{tbl:una_periods_other_controlcountry}
  \begin{tabular}{p{3cm} p{1.5cm} p{2.5cm} p{1.5cm}p{4cm}}
    \multicolumn{5}{c}{Panel A: Hungary and Slovakia} \\
    \toprule
& Effect & Standard error & P-value& Number of observations \\
    \midrule
    \multicolumn{1}{l}{Frequency of sodas} & \multicolumn{1}{c}{0.06} & \multicolumn{1}{c}{0.09} & \multicolumn{1}{c}{ 0.51 } & \multicolumn{1}{c}{16,012}\\
      \multicolumn{1}{l}{Body Mass Index (BMI)} & \multicolumn{1}{c}{-0.33} & \multicolumn{1}{c}{0.14} & \multicolumn{1}{c}{0.02} & \multicolumn{1}{c}{14,866} \\
    \bottomrule
    \\
    \multicolumn{5}{c}{Panel B: France and Switzerland} \\
     \toprule
& Effect & Standard error & P-value& Number of observations \\
    \midrule
    \multicolumn{1}{l}{Frequency of sodas} & \multicolumn{1}{c}{-0.01} & \multicolumn{1}{c}{ 0.06} & \multicolumn{1}{c}{ 0.84} & \multicolumn{1}{c}{22,845} \\
    \multicolumn{1}{l}{Body Mass Index (BMI)} & \multicolumn{1}{c}{-0.13} & \multicolumn{1}{c}{0.09} & \multicolumn{1}{c}{ 0.15} & \multicolumn{1}{c}{ 20,646} \\
    \bottomrule
    \\
    \par

  \end{tabular}\\
  {\footnotesize Note: Standard errors are estimated by bootstrap.}
\end{table}


\newpage


\newpage

\begin{table}[!ht]
  \centering
  \caption{Unaffected periods with school-year specific clusters}
  \label{tbl:una_periods_other_controlcountry_school-year specific clusters}
   \begin{tabular}{p{3cm} p{1.5cm} p{2.5cm} p{1.5cm}p{4cm}}
    
    \multicolumn{5}{c}{Panel B: France and Switzerland} \\
     \toprule
& Effect & Standard error & P-value& Number of observations \\
    \midrule
    \multicolumn{1}{l}{Frequency of sodas} & \multicolumn{1}{c}{-0.01} & \multicolumn{1}{c}{0.07 } & \multicolumn{1}{c}{ 0.86} & \multicolumn{1}{c}{22,845} \\
    \multicolumn{1}{l}{Body Mass Index (BMI)} & \multicolumn{1}{c}{-0.13} & \multicolumn{1}{c}{0.11} & \multicolumn{1}{c}{0.21 } & \multicolumn{1}{c}{20,646 } \\
    \bottomrule
    \\
    \par

  \end{tabular}\\
  {\footnotesize Note: Bootstrapped and clustered standard errors.}
\end{table}


\vskip 3 cm
%

\begin{table}[!ht]
  \centering
  \caption{Results with school-year specific clusters}
  \label{tbl:results_other_controlcountry_school-year specific clusters}
   \begin{tabular}{p{3cm} p{1.5cm} p{2.5cm} p{1.5cm}p{4cm}}
    \multicolumn{5}{c}{Panel A: Hungary and Slovakia} \\
    \toprule
& Effect & Standard error & P-value& Number of observations \\
    \midrule
    \multicolumn{1}{l}{Frequency of sodas} & \multicolumn{1}{c}{0.48} & \multicolumn{1}{c}{0.11} & \multicolumn{1}{c}{0.00} & \multicolumn{1}{c}{15,425}\\
    \multicolumn{1}{l}{Body Mass Index (BMI)} & \multicolumn{1}{c}{-0.13} & \multicolumn{1}{c}{0.18} & \multicolumn{1}{c}{0.46}& \multicolumn{1}{c}{14,059}\\
    \bottomrule
     \\
    \multicolumn{5}{c}{Panel B: France and Switzerland} \\
     \toprule
& Effect & Standard error & P-value& Number of observations \\
    \midrule
    \multicolumn{1}{l}{Frequency of sodas} & \multicolumn{1}{c}{-0.04} & \multicolumn{1}{c}{0.07} & \multicolumn{1}{c}{0.57} & \multicolumn{1}{c}{22,986}\\
    \multicolumn{1}{l}{Body Mass Index (BMI)} & \multicolumn{1}{c}{-0.13} & \multicolumn{1}{c}{0.11} & \multicolumn{1}{c}{0.24} & \multicolumn{1}{c}{20,258} \\
    \bottomrule
    \\
    \par

  \end{tabular}\\
  {\footnotesize Note: Bootstrapped and clustered standard errors.}
\end{table}

\newpage

\begin{figure}[h!]
\centering
\includegraphics[width=0.75\textwidth]{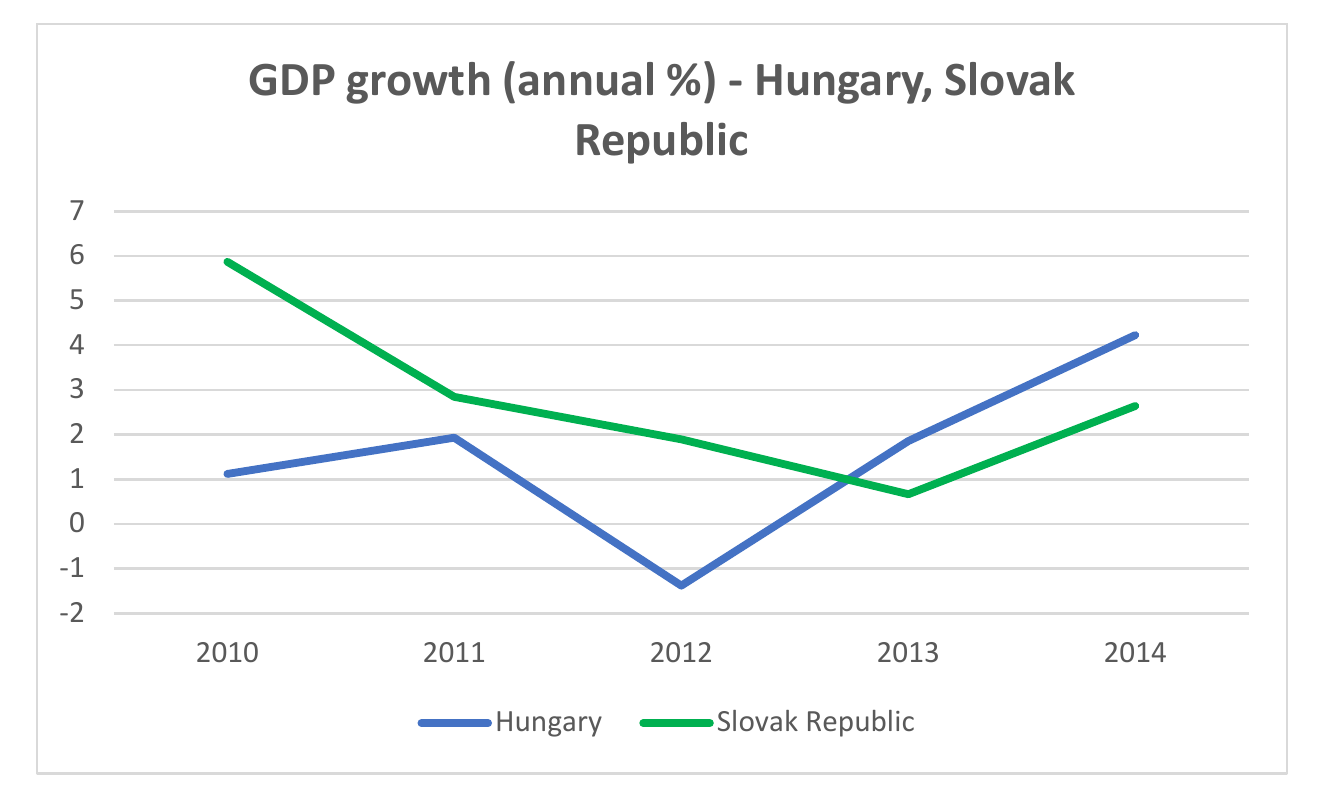}\\
\caption{Parallel GDP growth in Hungary and Slovakia}
\label{fig: parallel GDP growth in H and Slo}
{\footnotesize Source: The World Bank\footnotemark}
\end{figure}

\footnotetext{\url{https://data.worldbank.org/indicator/NY.GDP.MKTP.KD.ZG?end=2014&locations=HU-SK&start=2010}, last accessed: 29.03.2021.}

\vskip 2 cm

\begin{figure}[h!]
\centering
\includegraphics[width=0.75\textwidth]{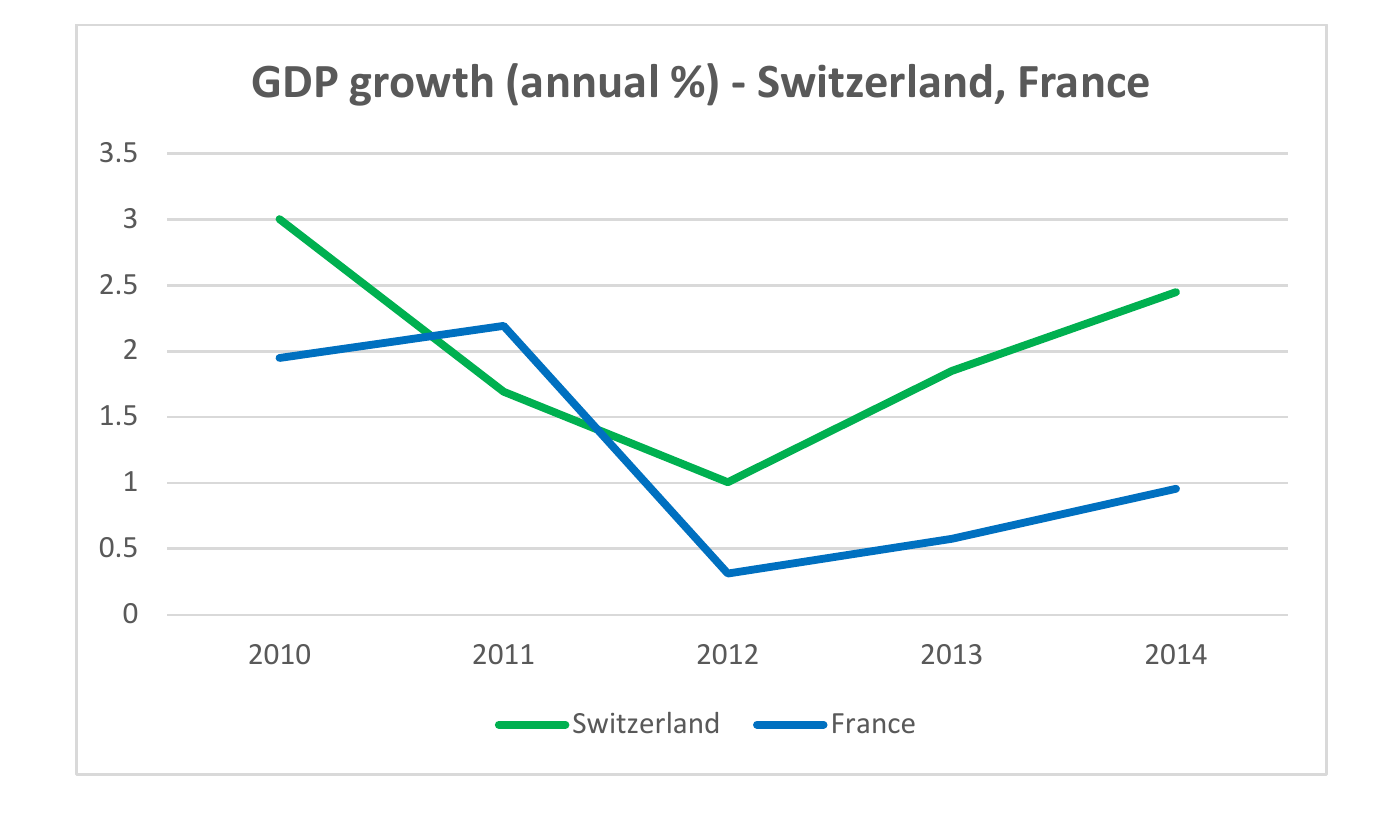}\\
\caption{Parallel GDP growth in France and Switzerland}
\label{GDPgrowthFSw}
{\footnotesize Source: The World Bank\footnotemark}

\end{figure}

\footnotetext{\url{https://data.worldbank.org/indicator/NY.GDP.MKTP.KD.ZG?end=2014&locations=HU-SK&start=2010}, last accessed: 29.03.2021.}

\end{appendices}
\newpage

\begingroup
    \setlength{\bibsep}{10pt}
    \setstretch{1}
   \bibliography{DissResearch}{}
\endgroup

\end{document}